\documentclass{article}
\usepackage{iclr2027_conference,times}
\usepackage{graphicx}
\usepackage{float}
\usepackage{wrapfig}
\usepackage{amsmath,amssymb}
\usepackage{booktabs}
\usepackage{hyperref}
\usepackage{url}

\title{OT-FlowRec: Optimal Transport-Guided Multi-Bridge Flow for Multimodal Sequential Recommendation}
\author{Miao Sun \\
    Nanjing University of Science and Technology
School of Computer Science and Engineering \\
    \texttt{miao.sun@njust.edu.cn}
}

\iclrfinalcopy

\begin{document}

\maketitle

\fancyhead[L]{}

\begin{abstract}
  E-commerce user preferences evolve dynamically within sessions, exhibiting concurrent and multi-centric interest distributions. While existing multimodal sequential recommendation methods leverage global attention to compress historical interactions, they inadequately capture the concurrent nature of multi-interest behaviors and fail to model the underlying multi-centric distribution of user intent. To tackle this issue, we propose OT-FlowRec, an optimal transport-guided multi-bridge flow matching framework that explicitly models multi-interest distributions to facilitate multimodal sequential recommendation. Unlike single-path generative models, OT-FlowRec constructs a mixture of parallel, interest-specific transport flows to model the multi-centric distribution characteristics, where each flow preserves a distinct user interest cluster via an optimal transport plan. Then, a geometric transport mechanism is designed to establish minimum-cost paths from the latent multi-interest distribution to target preference prototypes to provide multi-interest awareness. Moreover, a progressive time-sampling warmup strategy dynamically biases optimization weights toward the final states of the generation process, bridging the gap between continuous global distribution fitting and practical top-$k$ ranking metrics. This approach achieves explicit interest disentanglement, ensuring the preservation of a multi-peaked preference representation instead of collapsing it into a unimodal average. Extensive experiments on three public datasets demonstrate that OT-FlowRec outperforms existing state-of-the-art baselines.
\end{abstract}

\section{Introduction}

Multimodal sequential recommendation (MSR) has emerged as a pivotal paradigm for delivering personalized services in e-commerce, driven by its capacity to model the temporal dynamics of user preferences. By synergistically integrating interaction sequences with complementary signals derived from text and images, MSR is advantageous in disambiguating items that share similar behavioral footprints yet differ semantically, meanwhile enriching preference inference from sparse or inherently noisy user interactions. This enables MSR to capture users' nuanced, often concurrent interests—a capability that pure sequential models lack, and has drawn growing research attention.
\citep{ref38_pan2026_aso,ref39_hu2023_amm,ref41_liu2025_mrs,ref43_ji2023_ode,ref45_liu2024_aaa,ref46_xu2025_ccg,ref49_xv2024_imm,ref50_wei2023_mms,ref51_zhang2025_mpt,ref52_liu2024_mpa,ref53_fang2024_mrs}.

A substantial body of research has been devoted to MSR, which can be categorized into three technical lines: traditional sequential methods, diffusion-based methods, and flow-based methods. Specifically, traditional sequential methods use recurrent networks or self-attention to capture ordered dependencies through discriminative learning \citep{ref26_hidasi2016_sbr,ref25_kang2018_sas,ref27_sun2019_bsr}. Diffusion-based methods learn stochastic denoising processes in continuous spaces, improving robustness to noisy interactions and supporting flexible generation of target-item representations \citep{ref01_xie2024_bud,ref11_li2024_dad,ref13_yang2023_gwy,ref14_liu2025_pdf}. Flow-based methods employ vector-field learning and deterministic ODE evolution, reducing the stochasticity and iterative cost of diffusion sampling while directly modeling continuous preference trajectories \citep{ref18_liu2025_fmb,ref19_liu2025_fmf,ref20_huang2025_fmf,ref21_li2025_ptm}.

\begin{wrapfigure}{R}{0.55\textwidth}
  \centering
  \includegraphics[width=\linewidth]{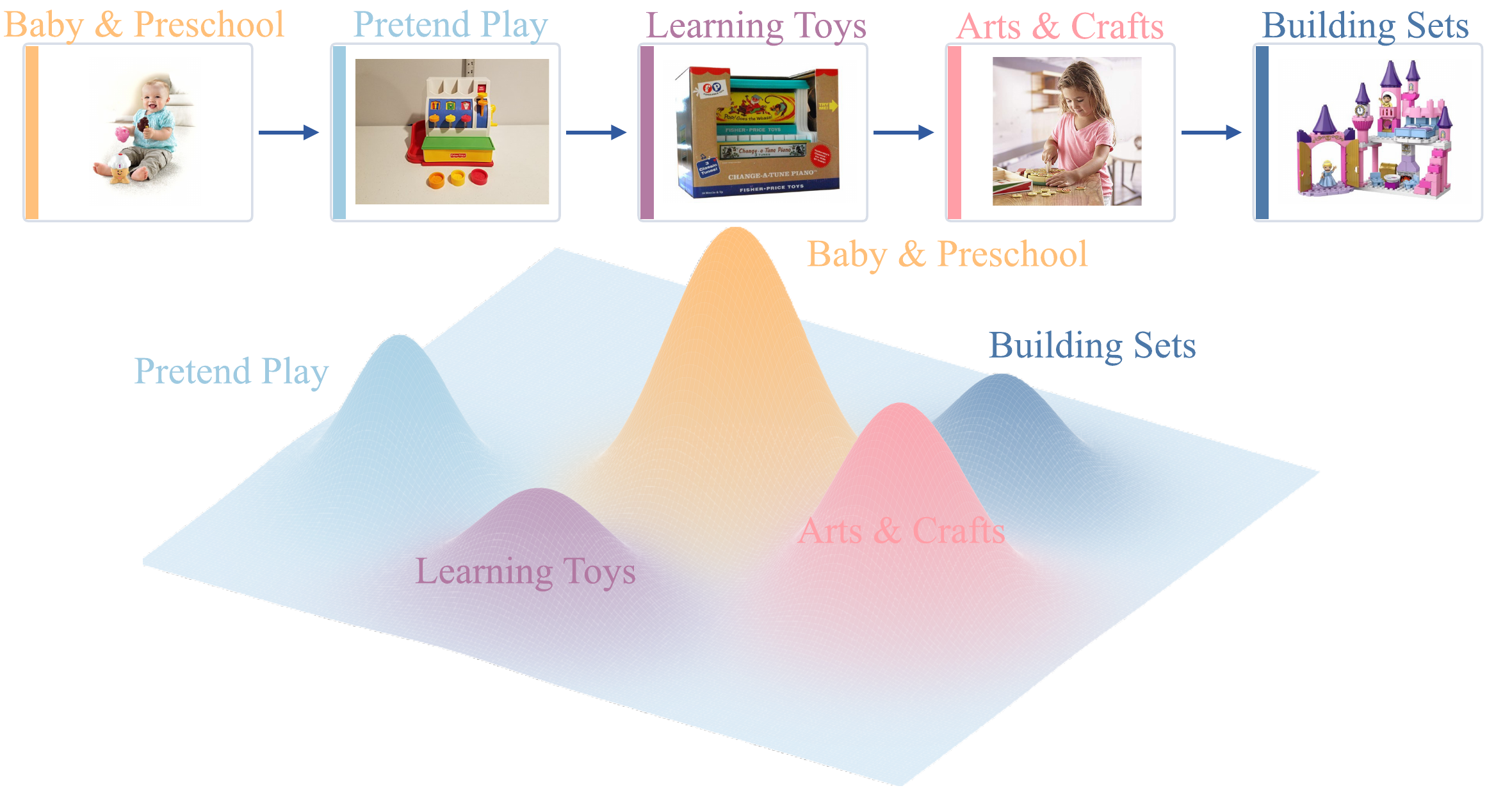}
  \caption{User interaction paths in the Toys dataset. Colors denote latent interest categories; this shows that multiple interests coexist within a session.}
  \label{fig:1}
\end{wrapfigure}
Although promising success has been achieved by previous works, two primary limitations remain. (i) \textbf{Multi-centric interest distribution collapse.} Existing global attention-based approaches compress multi-peaked interactions into single deterministic embeddings, which fail to describe the intrinsic multi-centric distribution characteristics. Moreover, most existing diffusion-based and flow-based methods construct single-peaked distribution representation, which is insufficient to model the multi-centric distribution of user intent.~\citep{ref25_kang2018_sas,ref27_sun2019_bsr,ref13_yang2023_gwy,ref08_hu2024_gai,ref18_liu2025_fmb,ref21_li2025_ptm} (Figure~\ref{fig:1}). (ii) \textbf{Static transport with misaligned priors.} Prevailing flow-based generative recommenders construct time-dependent conditional paths based on independently sampled prior distributions, without accounting for the underlying optimal transport geometry between users' historical interactions and their target preferences. Consequently, the learned trajectories are directionally unconstrained and lack semantic grounding to the multi-interest prototypes \citep{ref54_lipman2023_fmf,ref55_tong2024_iag,ref19_liu2025_fmf}.

To address these limitations, we propose OT-FlowRec, an optimal transport-guided multi-bridge flow matching framework.
Rather than collapsing a complex interaction sequence into a single representation, a Multi-Prototype Guided Source Distribution (MPSD) module is constructed to represent concurrent user preferences with multi-interest source distribution.
Its interest-specific components initialize parallel transport flows, preserving the underlying multi-peaked structure and avoiding unimodal collapse.
Considering the advantage of interest prototypes~\citep{wang2023missrec} in capturing diverse user semantics, an OT-Guided Multi-Bridge Flow Matching (OT-MBFM) module is further built, where multiple preference prototypes are first constructed. Accordingly, an interest distribution-to-prototype transformation learning process is then designed to map the multiple interest distribution to their matched prototypes through those parallel transport flows.
To facilitate the transformation, minimum-cost paths are computed to connect the latent source distribution to target preference prototypes, ensuring explicit awareness of diverse user interests during generation.
Furthermore, a mismatch remains between continuous distribution fitting and discrete ranking evaluation during the sampling process. We therefore develop a Progressive Time-sampling Warmup (PTW) that progressively shifts optimization toward the terminal states of the generation trajectory.
By focusing training on these later states, PTW narrows the gap between continuous trajectory fitting and top-$k$ ranking.
The main contributions are summarized as follows:
\begin{itemize}
  \item We propose OT-FlowRec, a multi-bridge flow learning framework that explicitly encodes the intrinsic multi-centric distribution of user interests. Unlike global attention mechanisms that compress concurrent interests into single deterministic embeddings, OT-FlowRec constructs a mixture of parallel, interest-specific transport flows, achieving effective multi-interest disentanglement while avoiding the multi-centric interest distribution collapse commonly observed in global attention methods.
  \item We propose the OT-MBFM module which establishes minimum-cost paths from the latent multi-interest distribution to target preference prototypes, providing directionally constrained trajectories. Specifically, a PTW strategy is designed to improve ranking quality by dynamically shifting optimization weights towards the terminal generative states, which effectively aligns continuous distribution fitting with top-$k$ ranking metrics.
  \item Extensive experiments on three public datasets demonstrate that our framework outperforms existing state-of-the-art baselines across all evaluation metrics.
\end{itemize}

\section{Methodology}
\subsection{Preliminaries}

In sequential multimodal recommendation, the user set is defined as $U=\{u_{1},u_{2},\ldots ,u_{|U|}\}$ and the item set is defined as $I=\{i_{1},i_{2},\ldots ,i_{|I|}\}$. For each user $u\in U$, the sequence of historically interacted items is represented by $S_{u}=[i_{1}^{(u)},i_{2}^{(u)},\ldots ,i_{n}^{(u)}]$, where $n$ denotes sequence length.

\subsection{Overall Framework}

\begin{figure}[!tbp]
  \centering
  \includegraphics[width=0.92\linewidth]{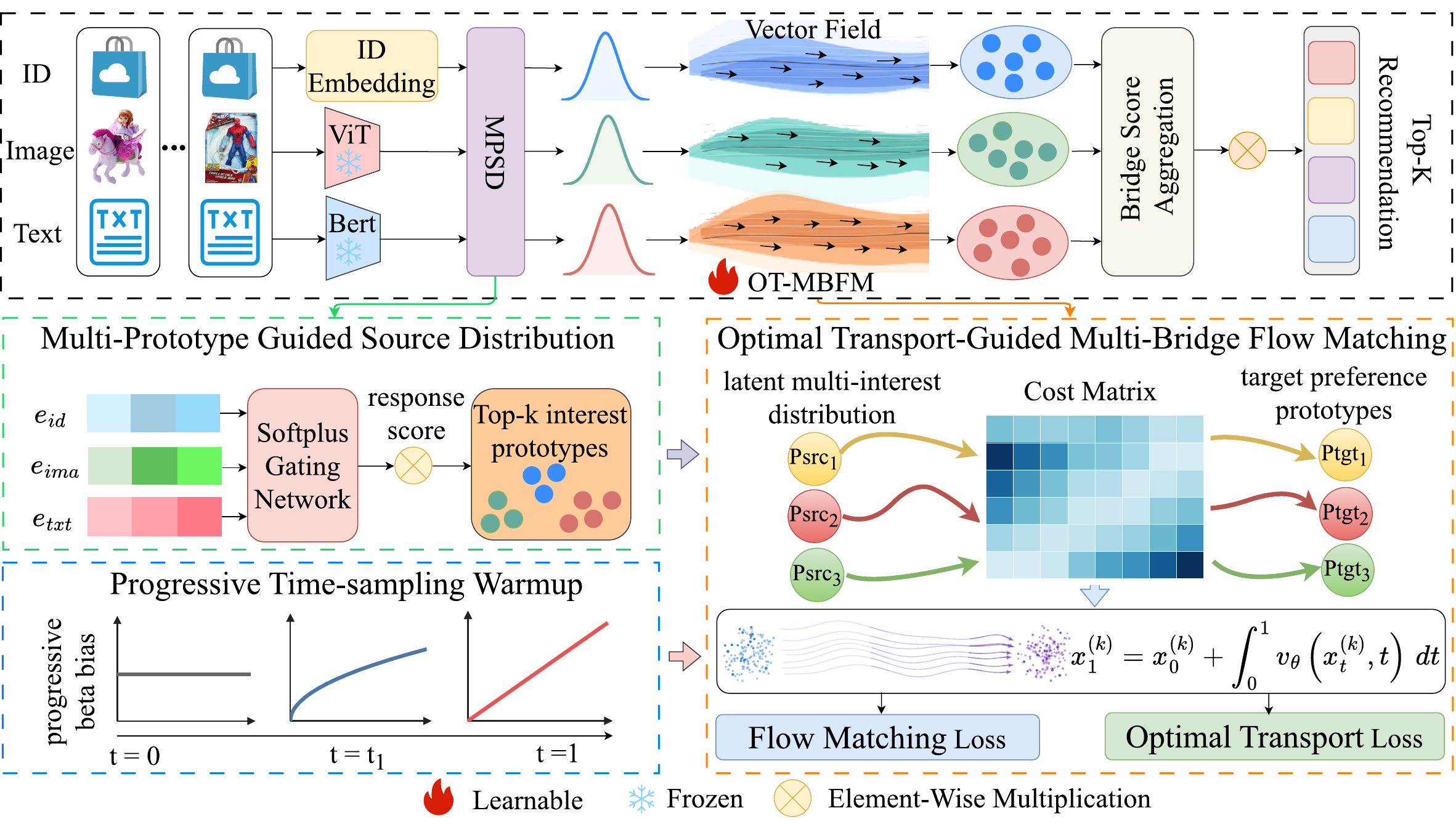}
  \caption{The overall architecture of OT-FlowRec. Top: The end-to-end recommendation pipeline, transforming multimodal item interactions into parallel transport flows. Middle-left: MPSD fuses multimodal features via a softplus gating network and extracts top-$k$ prototypes to construct parallel interest distributions. Bottom-right: OT-MBFM establishes OT couplings via a geometric cost matrix to guide deterministic flow matching along minimum-cost trajectories. Bottom-left: PTW dynamically shifts time-step sampling weights toward terminal generation states ($t \to 1$) via Beta distribution adjustments.}
  \label{fig:2}
\end{figure}

Figure~\ref{fig:2} shows OT-FlowRec coordinating three modules to transform multimodal interaction sequences into top-$k$ recommendation lists.

Given multimodal interaction sequences, OT-FlowRec fuses item ID embeddings with pre-trained visual and textual features via a softplus gating network. Matching these representations against a learnable prototype space selects top-$k$ responsive prototypes to initialize parallel interest-specific bridges. An optimal transport (OT) matrix couples these bridges with target preference prototypes along minimum-cost trajectories, letting a $v_\theta$-guided ODE solver integrate states over $t \in [0,1]$ into $k$ terminal representations. During training, a dynamic Beta distribution replaces uniform sampling, shifting weights toward late intervals to align trajectory fitting with discrete ranking objectives. At inference, a bridge score aggregation layer max-pools inner products between terminal states and candidate item embeddings for top-$k$ recommendations.


\subsection{Multi-Prototype Guided Source Distribution}

We propose the MPSD module to explicitly disentangle concurrent user preferences. Rather than compressing multimodal interactions into a single representation via global attention \citep{ref25_kang2018_sas,ref27_sun2019_bsr,ref28_lin2024_msa}, MPSD evaluates user responsiveness to multi-interest prototypes. By selecting top-$k$ prototypes to construct independent ODE paths, it forms parallel transport flows that preserve multi-peaked distribution of next-item preferences.

To exploit the multimodal semantics of items, a softplus gating network is designed to dynamically fuse collaborative filtering signals and modal features. For any item $i$ in the sequence and modality $m\in\{id,v,t\}$, its multimodal fused representation $x_{i}\in \mathbb{R}^{d}$ is calculated as follows:

\begin{equation}
  g_{m}=\operatorname{Softplus}\left(W_{m}\left[e_{id}\Vert e_{v}\Vert e_{t}\right]+b_{m}\right),
\end{equation}

\begin{equation}
  x_{i}=g_{id}\odot e_{id}+g_{v}\odot e_{v}+g_{t}\odot e_{t},
\end{equation}
where $W_{m}\in \mathbb{R}^{d\times 3d}$ and $b_{m}\in \mathbb{R}^{d}$ are learnable mapping parameters. This gating mechanism adaptively weights different modalities to accurately capture item characteristics.

To disentangle multiple user interests, a globally learnable interest prototype space $P=[p_{1},p_{2},\ldots ,p_{M}]\in \mathbb{R}^{M\times d}$ is constructed, where $M$ is the total number of prototypes. The model extracts the most recent fine-grained preference state $h_{u}\in \mathbb{R}^{d}$, which combines the last hidden state with the sequence representation and calculates its normalized response score $s_{j}$ with respect to all candidate prototypes $p_{j}$ as follows:

\begin{equation}
  s_{j}=\frac{\exp(h_{u}^{\top}p_{j}/\tau )}{\sum_{m=1}^{M}\,\exp(h_{u}^{\top}p_{m}/\tau )},
\end{equation}
where $\tau$ is the temperature parameter. This score quantifies the alignment between the user state and latent interests for subsequent prototype selection.

We extract the top-$k$ prototypes with the highest response scores instead of a global weighted average to preserve the multi-peaked nature of diverse user interests. The index set of the selected prototypes $K_{p}$ is defined as follows:

\begin{equation}
  K_{p}=\operatorname{TopK}\left(s_{1},\ldots ,s_{M}\right).
\end{equation}

Consequently, the representation strategy is transformed into a mixture of parallel, interest-specific transport flows. This design achieves interest disentanglement, thereby maintaining a robust multi-peaked preference representation.

\subsection{OT-Guided Multi-Bridge Flow Matching}

This section details the OT-MBFM module, which constructs deterministic transport trajectories from the latent multi-interest distribution to target preference prototypes. It minimizes geometric transport cost to derive an optimal coupling that guides continuous representation generation.

Assume that the selected top-$k$ interest prototype matrix $P_{tgt}\in \mathbb{R}^{|K_{p}|\times d}$ serves as the support set for the target distribution. The geometric transport cost matrix $C\in \mathbb{R}^{n\times |K_{p}|}$ between the latent multi-interest distribution $P_{src}$ and $P_{tgt}$ is formulated as $C_{j,k}=\Vert x_{j}-p_{k}\Vert_{2}^{2}$. The marginal masses are specified in Appendix~\ref{app:ot_marginals}. By introducing Sinkhorn entropy regularization \citep{ref63_cuturi2013_sdl}, the OT plan $\pi^{\star}\in \mathbb{R}^{n\times |K_{p}|}$ is obtained as follows:

\begin{equation}
  \pi^{\star}=\arg\min_{\pi \in\Pi(P_{src},P_{tgt})}\,\sum_{j,k}\,\pi_{j,k}C_{j,k}-\epsilon H(\pi ),
\end{equation}

\begin{equation}
  H(\pi )=-\sum_{j,k}\,\pi_{j,k}\log\pi_{j,k},
\end{equation}
where $\Pi(\cdot ,\cdot )$ is the set of all joint probability measures that satisfy the marginal distribution constraints, and $\epsilon$ denotes a hyperparameter controlling smoothness.

Based on $\pi^{\star}$, the framework softly assigns historical sequence items to corresponding target preference prototypes, thereby forming multiple parallel flow matching bridges. For any time-step $t \in [0, 1]$, a deep neural network $v_{\theta}(x,t)$ parameterizes the target vector field $v_{t}(x)$. Guided by the OT plan, the conditional vector field transitions from random pairing to deterministic matching along optimal geometric paths. By establishing minimum-cost transport paths, the OT-MBFM module equips the framework with explicit multi-interest awareness throughout continuous generation.

\subsection{Progressive Time-sampling Warmup}

This section details the PTW strategy, designed to align continuous generative objectives with discrete top-$k$ ranking metrics. By dynamically shifting optimization weights toward the terminal states of the generation trajectory, the PTW strategy ensures that the vector field precision is concentrated where practical recommendation ordering is determined.

Specifically, the strategy employs the Beta distribution to replace the standard uniform distribution $U(0,1)$ for sampling the integration time-step $t$. The shape parameter $\alpha$ is dynamically adjusted as training progresses:

\begin{equation}
  t\sim \operatorname{Beta}\left(\alpha ,1\right).
\end{equation}

This formulation establishes a time-dependent weighting mechanism to prioritize late-stage generation. During the early stages of training, setting $\alpha \approx 1$ reduces the process to uniform sampling, which prompts the model to capture the global vector field. As training advances, $\alpha$ is gradually increased to bias the sampling probability toward the high time-step interval approaching $1$. Consequently, this mechanism transitions the optimization focus from global distribution modeling to terminal state refinement, bridging continuous distribution fitting and discrete ranking evaluation.

\subsection{Optimization}

This section details the end-to-end optimization of OT-FlowRec. To guide continuous representation generation and discrete item ranking, the framework minimizes a combined loss function $L_{\text{total}}$:
\begin{equation}
  \mathcal{L}_{\mathrm{total}} = \lambda_{1}\mathcal{L}_{\mathrm{FM}} + \lambda_{2}\mathcal{L}_{\mathrm{CE}} + \lambda_{3}\mathcal{L}_{\mathrm{OT}} + \lambda_{4}\mathcal{L}_{\mathrm{Proto}},
\end{equation}
where $\mathcal{L}_{\mathrm{FM}}$, $\mathcal{L}_{\mathrm{CE}}$, $\mathcal{L}_{\mathrm{OT}}$, and $\mathcal{L}_{\mathrm{Proto}}$ denote the flow matching loss, the cross-entropy ranking loss, the OT geometric regularization, and the prototype orthogonality constraint, respectively, with $\lambda_{\cdot}$ representing weighting hyperparameters tuned via Bayesian optimization.

To drive representation generation along the deterministic geometric paths established by the OT-MBFM module, the target vector field $v_{t}(x)$ at time-step $t \in [0, 1]$ is parameterized by a deep neural network $v_{\theta}(x,t)$. The flow matching loss $\mathcal{L}_{\mathrm{FM}}$ is defined as follows:
\begin{equation}
  \mathcal{L}_{\mathrm{FM}}(\theta)
  =
  \mathbb{E}_{\substack{
        t\sim\operatorname{Beta}(\alpha,1),\\
        (x_0,x_1)\sim\pi^\star
      }}
  \left[
    \left\|
    v_\theta(x_t,t)-(x_1-x_0)
    \right\|_2^2
    \right],
\end{equation}
where $x_{t}=(1-t)x_{0}+tx_{1}$ represents interpolated state along OT path. This loss guides continuous representation generation along minimum-cost transport paths.

Bridging continuous generation and discrete ranking, $\mathcal{L}_{\mathrm{CE}}$ injects a collaborative filtering signal by aggregating the terminal states $x_1^{(k)}$ over the selected prototypes $K_p$. Specifically, we define the bridge score for item $j$ as
$s_{j}=\max_{k\in K_p}
  \langle x_{1}^{(k)},e_j\rangle/\tau_{\mathrm{ce}}$
and maximize the likelihood of the ground-truth next item
$i_{n+1}$:
\begin{equation}
  \mathcal{L}_{\mathrm{CE}}
  =
  -\sum_{u\in\mathcal{U}}
  \log
  \frac{
    \exp\left(s_{i_{n+1}}\right)
  }{
    \sum_{j\in\mathcal{I}}\exp\left(s_{j}\right)
  },
\end{equation}
where $\tau_{ce}$ is the temperature and $e$ denotes item embeddings.

To prevent unconstrained generation in semantic spaces, we regularize vector field using the Sinkhorn optimal coupling $\pi^\star$ between source distribution and target prototypes for each bridge:
\begin{equation}
  \mathcal{L}_{\mathrm{OT}}
  =
  \sum_{j=1}^{n}
  \sum_{k\in K_p}
  \pi_{j,k}^{\star}
  \left\lVert x_j-p_k \right\rVert_2^2
  -
  \epsilon H\left(\pi^{\star}\right),
  \label{OT}
\end{equation}
with entropy regularizer $H(\cdot)$. This regularization guides generation trajectories along minimal-effort transport paths.

Finally, the orthogonality constraint $\mathcal{L}_{\mathrm{Proto}}$ on the row-normalized matrix $\tilde{P}$ maintains semantic capacity against mode collapse:
\begin{equation}
  \mathcal{L}_{\mathrm{Proto}} = \|\tilde{P}\tilde{P}^\top - I_M\|_F^2.
\end{equation}
This constraint disperses prototypes uniformly to capture multi-peaked interest independently.

We optimize the framework using Adam. During inference, the selected branches share the initial state $x_0^{(k)}=h_{u}$, the Euler ODE solver integrates along the learned vector field:
\begin{equation}
  x_1^{(k)} = x_0^{(k)} + \int_0^1 v_\theta(x_t^{(k)}, t\mid p_k) \, dt, \quad \forall k \in K_p,
\end{equation}

where the resulting terminal states $x_{1}^{(k)}$ serve as the predicted multi-interest representation. Finally, these representations are matched against candidate item embeddings via max-pooling aggregation to generate the top-$k$ recommendations.

\section{Experiments}

\subsection{Experiment Setting}

To evaluate the effectiveness of the proposed OT-FlowRec in sequential multimodal recommendation tasks, we design and conduct comprehensive experiments. Specifically, the experiments are guided by three primary research questions:

RQ1: We investigate whether OT-FlowRec consistently outperforms existing sequential, multimodal, and generative recommendation baselines across real-world datasets characterized by varying sequence lengths and item domains.

RQ2: We explore how the key modules of OT-FlowRec, including MPSD, OT-MBFM, and PTW, contribute to the overall recommendation performance, and we analyze the specific roles they play within the model.

RQ3: We examine how the proposed MPSD, OT-MBFM, and PTW modules improve recommendation performance and generation quality compared with traditional generative recommendation models that rely on single global representations and standard flow matching.

\subsubsection{Datasets and Evaluation Metrics}

\begin{wraptable}{r}{0.63\textwidth}
  \caption{Statistics of the evaluation datasets.}
  \label{tab:1}
  \centering
  {%
    \setlength{\tabcolsep}{3pt}
    \begin{tabular}{@{}lccccc@{}}
      \toprule
      Dataset     & Users & Items & Inter. & Sparsity & Avg. Len. \\
      \midrule
      Beauty      & 22364 & 12102 & 198502 & 99.92\%  & 8.8       \\
      Video Games & 24304 & 10679 & 231780 & 99.91\%  & 9.5       \\
      Toys        & 19413 & 11925 & 167597 & 99.92\%  & 8.6       \\
      \bottomrule
    \end{tabular}
  }
\end{wraptable}
We conduct experiments on three widely used Amazon datasets \citep{ref61_mcauley2015_ibr}: Beauty, Video Games, and Toys (statistics in Table~\ref{tab:1}). These datasets cover distinct item domains, representing cosmetics and personal care, gaming consoles and accessories, and children's entertainment products. For a fair comparison, we adopt Hit Rate (H@$K$) and Normalized Discounted Cumulative Gain (N@$K$) with $K \in \{10, 20\}$ to assess top-$k$ recall and ranking quality, respectively.

\subsubsection{Baseline Methods}

To evaluate OT-FlowRec, we compare it against three categories of baseline methods: (1) \textbf{Traditional sequential models}, including RNN-based and Transformer-based architectures (GRU4Rec \citep{ref26_hidasi2016_sbr}, SASRec \citep{ref25_kang2018_sas}, BERT4Rec \citep{ref27_sun2019_bsr}) and those integrating side information (MSSR \citep{ref28_lin2024_msa}); (2) \textbf{Diffusion-based models}, spanning foundational frameworks (DreamRec \citep{ref13_yang2023_gwy}, DiffuRec \citep{ref11_li2024_dad}) and advanced variants (DiffMM \citep{ref05_jiang2024_dmm}, ADRec \citep{ref03_chen2025_utp}, MoDiCF \citep{ref09_li2025_gwf}, PreferDiff \citep{ref14_liu2025_pdf}); and (3) \textbf{Flow-based models} tailored for recommendation (FMRec \citep{ref18_liu2025_fmb}, FlowCF \citep{ref19_liu2025_fmf}, FindRec \citep{ref22_wang2025_fsg}).

\subsubsection{Implementation Details}

We implement the proposed method using the widely adopted RecBole \citep{ref62_zhao2021_rta} framework. For all datasets, we filter out users and items with fewer than 5 interactions and truncate historical sequences to a maximum length of 50. The models are trained for up to 300 epochs using the Adam optimizer, with a training batch size of 1024 and an evaluation batch size of 4096. We apply an early-stopping strategy if validation performance does not improve for 5 consecutive evaluations. Performance is evaluated using Hit and NDCG at $K \in \{10, 20\}$. Textual and visual features are extracted using BERT \citep{ref60_devlin2019_bpt} (768 dimensions) and CLIP-ViT \citep{ref59_radford2021_ltv} (512 dimensions), respectively, which are then mapped to the hidden dimension $d$ via linear layers. All experiments are conducted on a single NVIDIA RTX A6000 GPU. For further experimental details, see Appendix~\ref{app:additional_implementation_details}.

\subsection{RQ1: Overall Performance}

Table~\ref{tab:2} demonstrates that OT-FlowRec outperforms all three baseline categories across all datasets and top-$k$ metrics, achieving average performance improvements of 25.08\% in H@20 and 9.15\% in N@20 across the three datasets.

\textbf{Comparison with traditional sequential models.} Traditional sequential baselines compress complex historical interactions into a single deterministic representation via global attention, inducing unimodal collapse that prevents capturing concurrent multi-interest behaviors. Through the MPSD module, OT-FlowRec explicitly disentangles interactions into parallel, interest-specific transport flows to preserve multi-peaked interest distributions. Consequently, compared with MSSR, which is the strongest baseline in this category, OT-FlowRec achieves substantial improvements on Beauty dataset, increasing H@20 by 32.53\% and N@20 by 19.86\%.

\textbf{Comparison with diffusion-based models.} Continuous diffusion models suffer from unguided and curved diffusion paths. In contrast, OT-FlowRec replaces stochastic denoising with deterministic flow matching along OT paths. By establishing explicit minimum-cost connections between source distributions and target prototypes via the OT-MBFM module, the framework significantly stabilizes generation. Compared with ADRec, the top-performing diffusion baseline, OT-FlowRec demonstrates notable gains on Video Games, improving H@20 by 27.69\% and N@20 by 10.63\%.

\textbf{Comparison with flow-based models.} Existing flow-based recommendation baselines ignore semantic transport distances by relying on randomly paired Euclidean straight paths and uniform time sampling, which misaligns continuous trajectory fitting with discrete ranking objectives. OT-FlowRec resolves these bottlenecks through OT couplings and the PTW strategy. Biasing optimization weights toward terminal generation states ($t \to 1$) concentrates vector field precision where discrete item ranking occurs. Compared with FlowCF, the strongest flow-based competitor, OT-FlowRec drives NDCG improvements, increasing N@20 by 9.81\% over FlowCF on Beauty.

\begin{table}[!tbp]
  \caption{Overall performance. The best and second-best scores are shown in bold and underlined, respectively.}
  \label{tab:2}
  \centering
  \setlength{\tabcolsep}{1pt}
  \renewcommand{\arraystretch}{0.9}
  \resizebox{\textwidth}{!}{%
    \begin{tabular}{@{}p{1.75cm}*{12}{c}@{}}
      \toprule
      Baseline   & \multicolumn{4}{c}{Beauty} & \multicolumn{4}{c}{Video Games} & \multicolumn{4}{c}{Toys}                                                                                                                                                                                              \\
      \cmidrule(lr){2-5}\cmidrule(lr){6-9}\cmidrule(lr){10-13}
                 & H@10                       & H@20                            & N@10                     & N@20               & H@10               & H@20               & N@10               & N@20               & H@10               & H@20               & N@10               & N@20               \\
      \midrule
      SASRec     & 0.0401                     & 0.0660                          & 0.0200                   & 0.0265             & 0.0847             & 0.1281             & 0.0423             & 0.0532             & 0.0322             & 0.0497             & 0.0164             & 0.0208             \\
      GRU4Rec    & 0.0562                     & 0.0851                          & 0.0298                   & 0.0370             & 0.0725             & 0.1122             & 0.0376             & 0.0475             & 0.0454             & 0.0676             & 0.0260             & 0.0316             \\
      BERT4Rec   & 0.0379                     & 0.0571                          & 0.0200                   & 0.0249             & 0.0708             & 0.1091             & 0.0361             & 0.0456             & 0.0274             & 0.0395             & 0.0154             & 0.0184             \\
      MSSR       & 0.0583                     & 0.0956                          & 0.0314                   & 0.0448             & 0.0709             & 0.1283             & 0.0415             & 0.0515             & 0.0613             & 0.0950             & 0.0318             & 0.0410             \\
      \midrule
      DreamRec   & 0.0228                     & 0.0417                          & 0.0130                   & 0.0176             & 0.0415             & 0.0712             & 0.0221             & 0.0295             & 0.0216             & 0.0391             & 0.0111             & 0.0155             \\
      DiffuRec   & 0.0548                     & 0.0939                          & 0.0263                   & 0.0354             & 0.0617             & 0.0795             & 0.0278             & 0.0433             & 0.0725             & 0.0982             & 0.0366             & 0.0430             \\
      DiffMM     & 0.0556                     & 0.0975                          & 0.0225                   & 0.0361             & 0.0651             & 0.1064             & 0.0333             & 0.0438             & 0.0416             & 0.0664             & 0.0214             & 0.0276             \\
      ADRec      & \underline{0.0673}         & 0.0945                          & \underline{0.0417}       & 0.0485             & \underline{0.1059} & \underline{0.1538} & \underline{0.0584} & \underline{0.0705} & 0.0704             & 0.0909             & 0.0421             & \underline{0.0512} \\
      MoDiCF     & 0.0531                     & 0.0821                          & 0.0263                   & 0.0335             & 0.0585             & 0.0948             & 0.0292             & 0.0383             & 0.0496             & 0.0724             & 0.0254             & 0.0312             \\
      PreferDiff & 0.0594                     & 0.0675                          & 0.0350                   & 0.0471             & 0.0735             & 0.0838             & 0.0552             & 0.0578             & 0.0463             & 0.0489             & 0.0334             & 0.0341             \\
      \midrule
      FMRec      & 0.0531                     & 0.1069                          & 0.0358                   & 0.0459             & 0.0918             & 0.1370             & 0.0477             & 0.0593             & \underline{0.0838} & \underline{0.1013} & 0.0426             & 0.0501             \\
      FlowCF     & 0.0502                     & \underline{0.1075}              & 0.0349                   & \underline{0.0489} & 0.0744             & 0.1050             & 0.0427             & 0.0505             & 0.0650             & 0.0804             & 0.0360             & 0.0450             \\
      FindRec    & 0.0549                     & 0.0770                          & 0.0318                   & 0.0374             & 0.0911             & 0.1394             & 0.0465             & 0.0586             & 0.0703             & 0.0959             & \underline{0.0440} & 0.0505             \\
      \midrule
      ours       & \textbf{0.0891}            & \textbf{0.1267}                 & \textbf{0.0442}          & \textbf{0.0537}    & \textbf{0.1276}    & \textbf{0.1964}    & \textbf{0.0607}    & \textbf{0.0780}    & \textbf{0.0946}    & \textbf{0.1314}    & \textbf{0.0455}    & \textbf{0.0548}    \\
      Improv.    & 32.39\%                    & 17.86\%                         & 5.99\%                   & 9.81\%             & 20.49\%            & 27.69\%            & 3.93\%             & 10.63\%            & 12.88\%            & 29.71\%            & 3.40\%             & 7.03\%             \\
      \bottomrule
    \end{tabular}%
  }

  \par\medskip
  \caption{Ablation study results.}
  \label{tab:3}
  \centering
  \small
  \setlength{\tabcolsep}{1pt}
  \renewcommand{\arraystretch}{0.9}
  \resizebox{\textwidth}{!}{%
    \begin{tabular}{@{}p{2.0cm}*{12}{c}@{}}
      \toprule
      Variant     & \multicolumn{4}{c}{Beauty} & \multicolumn{4}{c}{Video Games} & \multicolumn{4}{c}{Toys}                                                                                                                                                                   \\
      \cmidrule(lr){2-5}\cmidrule(lr){6-9}\cmidrule(lr){10-13}
                  & H@10                       & H@20                            & N@10                     & N@20            & H@10            & H@20            & N@10            & N@20            & H@10            & H@20            & N@10            & N@20            \\
      \midrule
      w/o MPSD    & 0.0866                     & 0.1249                          & 0.0431                   & 0.0515          & 0.1241          & 0.1948          & 0.0590          & 0.0776          & 0.0934          & 0.1299          & 0.0437          & 0.0529          \\
      w/o OT-MBFM & 0.0874                     & 0.1253                          & 0.0435                   & 0.0531          & 0.1239          & 0.1936          & 0.0589          & 0.0764          & 0.0913          & 0.1303          & 0.0446          & 0.0544          \\
      w/o PTW     & 0.0863                     & 0.1257                          & 0.0428                   & 0.0527          & 0.1244          & 0.1919          & 0.0587          & 0.0757          & 0.0942          & 0.1299          & 0.0438          & 0.0529          \\
      ours        & \textbf{0.0891}            & \textbf{0.1267}                 & \textbf{0.0442}          & \textbf{0.0537} & \textbf{0.1276} & \textbf{0.1964} & \textbf{0.0607} & \textbf{0.0780} & \textbf{0.0946} & \textbf{0.1314} & \textbf{0.0455} & \textbf{0.0548} \\
      \bottomrule
    \end{tabular}%
  }
\end{table}


\subsection{RQ2: Ablation Study}

To assess the individual contributions of core designs in OT-FlowRec, we evaluate three ablation variants across the Beauty, Video Games, and Toys datasets (Table~\ref{tab:3}).

At the representation level, constructing parallel, interest-specific transport flows is core to disentangling concurrent user behaviors and preserving multi-peaked interest distributions. Removing the multi-prototype mechanism (\textbf{w/o MPSD}) compresses complex historical interactions into a single global representation, degrading performance across all evaluation metrics. For instance, on the Beauty dataset, H@10 drops from 0.0891 to 0.0866.

Moreover, establishing minimum-cost geometric paths from the latent multi-interest distribution to target preference prototypes contributes to explicit multi-interest awareness and mitigates trajectory conflicts. Replacing OT guidance with randomly paired Euclidean straight paths (\textbf{w/o OT-MBFM}) fails to capture complex behavioral sequences, causing performance declines. Specifically, H@10 decreases from 0.1276 to 0.1239 on the Video Games dataset.

During the optimization phase, dynamically shifting training weights toward terminal generation states is beneficial for continuous distribution fitting with discrete top-$k$ ranking metrics. Substituting the PTW strategy with uniform time-sampling (\textbf{w/o PTW}) ignores ranking sensitivity to late-stage integration precision ($t \to 1$). Consequently, this variant impairs ranking accuracy across all datasets, as evidenced by N@20 dropping from 0.0780 to 0.0757 on the Video Games dataset.


\subsection{RQ3: Qualitative Analysis}

\begin{wrapfigure}{R}{0.52\textwidth}
  \centering
  \includegraphics[width=\linewidth]{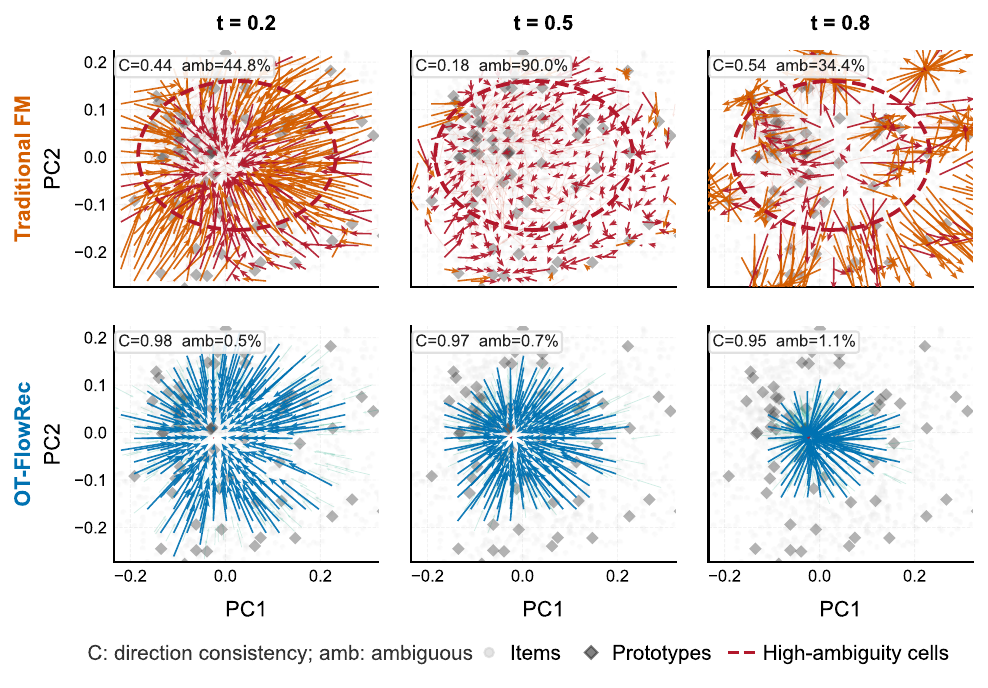}
  \caption{Vector field comparison between traditional flow matching and OT-FlowRec on the Toys.}
  \label{fig:5}
\end{wrapfigure}
Figure~\ref{fig:5} shows the vector field directions of both traditional random-pairing flow matching and OT-FlowRec at different integration times. According to Figure~\ref{fig:5}, traditional flow matching produces conflicting directions and extensive region ambiguity, with directional consistency falling to $0.18$ and the ambiguity rate rising to $90.0\%$ at $t=0.5$. In contrast, OT-FlowRec maintains directional consistency above $0.95$ and ambiguity below $1.1\%$ across all three time steps, resulting in more coherent transport trajectories. The comparison of vector field direction demonstrates that OT-guided coupling in OT-FlowRec provides directionally constrained trajectories.
The t-SNE and transport-cost analyses are presented in Appendix~\ref{app:qualitative_analysis}(Figures~\ref{fig:qualitative_tsne} and~\ref{fig:transport_costs}) and training stability
is presented in Appendix~\ref{app:additional_experimental_results}.

\section{Related Work}

\subsection{Traditional Sequential Recommendation}

Traditional sequential recommendation methods primarily learn behavior transition patterns from user historical interaction sequences \citep{ref30_cen2020_cmi,ref31_li2024_dit,ref32_zhang2024_dia,ref33_chen2022_icl,ref34_qin2024_icl,ref35_wang2019_mmp,ref36_li2019_min,ref37_tan2021_sin,ref38_pan2026_aso,ref44_park2025_tli,ref57_duan2025_llt,ref58_hou2022_tus}. Early approaches utilized Markov chains, RNNs \citep{ref26_hidasi2016_sbr}, and CNNs \citep{ref65_tang2018_ptn}, which were later broadly superseded by Transformer-based architectures like SASRec \citep{ref25_kang2018_sas} and BERT4Rec \citep{ref27_sun2019_bsr} to better capture sequence dependencies.

To alleviate interaction sparsity, subsequent research incorporated auxiliary multimodal information (e.g., text and images) \citep{ref39_hu2023_amm,ref40_wang2024_asi,ref41_liu2025_mrs,ref42_kim2025_dds,ref43_ji2023_ode,ref45_liu2024_aaa,ref46_xu2025_ccg,ref47_yang2025_ccd,ref48_zhang2025_hta,ref49_xv2024_imm,ref50_wei2023_mms,ref51_zhang2025_mpt,ref52_liu2024_mpa,ref53_fang2024_mrs}, utilizing techniques like multi-sequence attention \citep{ref28_lin2024_msa} and modality experts \citep{ref29_guo2025_mss}. However, these methods typically remain within the discriminative modeling paradigm, compressing complex user behaviors into deterministic, single-point representations. This limitation prevents them from fully capturing the multi-peaked distributional nature and generative evolution of dynamic user interests.

\subsection{Diffusion-Based Recommendation}

Diffusion-based methods shift recommendation from discriminative matching to generative distribution modeling \citep{ref01_xie2024_bud,ref03_chen2025_utp,ref04_cui2025_mmm,ref05_jiang2024_dmm,ref07_liu2023_daf,ref09_li2025_gwf,ref10_qu2025_iad,ref12_mao2025_oee,ref14_liu2025_pdf,ref16_wang2023_drm}. Early works like DiffuRec \citep{ref11_li2024_dad} and DreamRec \citep{ref13_yang2023_gwy} reconstruct target item representations via stochastic noise-addition and reverse denoising processes, thereby capturing diverse user interests as continuous distributions rather than fixed vectors.

Subsequent research has enhanced this paradigm by integrating multimodal user-item graphs \citep{ref05_jiang2024_dmm}, aligning generation with personalized ranking objectives \citep{ref14_liu2025_pdf}, and employing token-level diffusion to prevent embedding collapse \citep{ref03_chen2025_utp}. Despite these advancements, the inherent reliance on curved probability paths and multi-step stochastic sampling limits their inference efficiency and often leads to error accumulation during the generative process.

\subsection{Flow-Based Recommendation}

Flow-based models replace the stochastic denoising of diffusion models with deterministic continuous generation, transporting prior distributions to target distributions via time-dependent vector fields \citep{ref54_lipman2023_fmf,ref55_tong2024_iag}. Recent works have adapted flow matching for recommendation by designing discrete flows for implicit feedback \citep{ref19_liu2025_fmf}, utilizing straight trajectories to reduce preference perturbations \citep{ref18_liu2025_fmb}, and employing Stein-guided entropic flows to capture multi-peaked sequential preferences \citep{ref22_wang2025_fsg}.

Despite improving inference efficiency and stability over diffusion models, existing flow matching methods typically construct simple, global straight paths. This ignores the explicit geometric transport costs between historical distributions and latent multi-interest prototypes. Furthermore, their continuous distribution fitting objectives often misalign with discrete top-$k$ ranking metrics. To bridge these gaps, our OT-FlowRec proposes OT-MBFM for explicit interest disentanglement, alongside the PTW strategy to directly optimize late-stage generation for practical ranking accuracy.

\section{Conclusion}

In this paper, we proposed OT-FlowRec to jointly model concurrent user interests in multimodal sequential recommendation. To overcome unimodal collapse, our framework constructed parallel transport flows that disentangled diverse interests and preserved multi-peaked preference distributions. Integrating minimum-cost geometric transport paths with the progressive time-sampling warmup strategy achieved explicit multi-interest awareness and bridged continuous distribution fitting and discrete top-$k$ ranking metrics. Extensive experiments across three datasets demonstrated that OT-FlowRec outperformed state-of-the-art baselines.

\subsection*{AI use statement}

AI tools were used to polish the writing of this paper.
\bibliography{references}
\bibliographystyle{iclr2027_conference}

\clearpage
\appendix

\section{Marginal Distributions for Optimal Transport}
\label{app:ot_marginals}

For a user with $n$ valid Transformer states, let
$\ell=\min(n,L)$, where $L$ is the maximum source-window length.
The source and target distributions are
\begin{equation}
  P_{src}=\sum_{j=1}^{\ell}a_j\delta_{x_{n-\ell+j}},
  \qquad
  P_{tgt}=\sum_{k=1}^{M}s_k\delta_{p_k},
  \qquad
  a_j=
  \frac{\exp[-\gamma(\ell-j)]}
  {\sum_{q=1}^{\ell}\exp[-\gamma(\ell-q)]},
\end{equation}
where $\delta$ denotes a point mass, $\gamma$ controls recency
weighting, and $s_k$ is the normalized prototype response
defined in the main text. Padding positions are excluded.
The feasible couplings satisfy
\begin{equation}
  \Pi(P_{src},P_{tgt})
  =
  \{\pi\in\mathbb{R}_{+}^{\ell\times M}:
  \pi\mathbf{1}_{|K_p|}=\mathbf{a},\
  \pi^\top\mathbf{1}_{\ell}=\mathbf{s}\}.
\end{equation}

\section{Problem Formulation}

Given the multimodal context, each item $i\in I$ comprises an ID embedding $e_{id}\in \mathbb{R}^{d}$, a visual feature $e_{v}\in \mathbb{R}^{d}$ and a textual feature $e_{t}\in \mathbb{R}^{d}$, where $d$ represents the dimension of the latent feature space. Formally, given the historical interaction sequence $S_{u}$, the core task is to learn a continuous generative mapping function $f_{\theta}(\cdot )$ to predict the item $i_{n+1}^{(u)}\in I$ with the most likely to interact in the next time-step $n+1$.

\section{Qualitative Analysis (RQ3)}
\label{app:qualitative_analysis}

\begin{figure}[H]
  \centering
  \begin{minipage}[t]{0.56\linewidth}
    \centering
    \includegraphics[width=\linewidth]{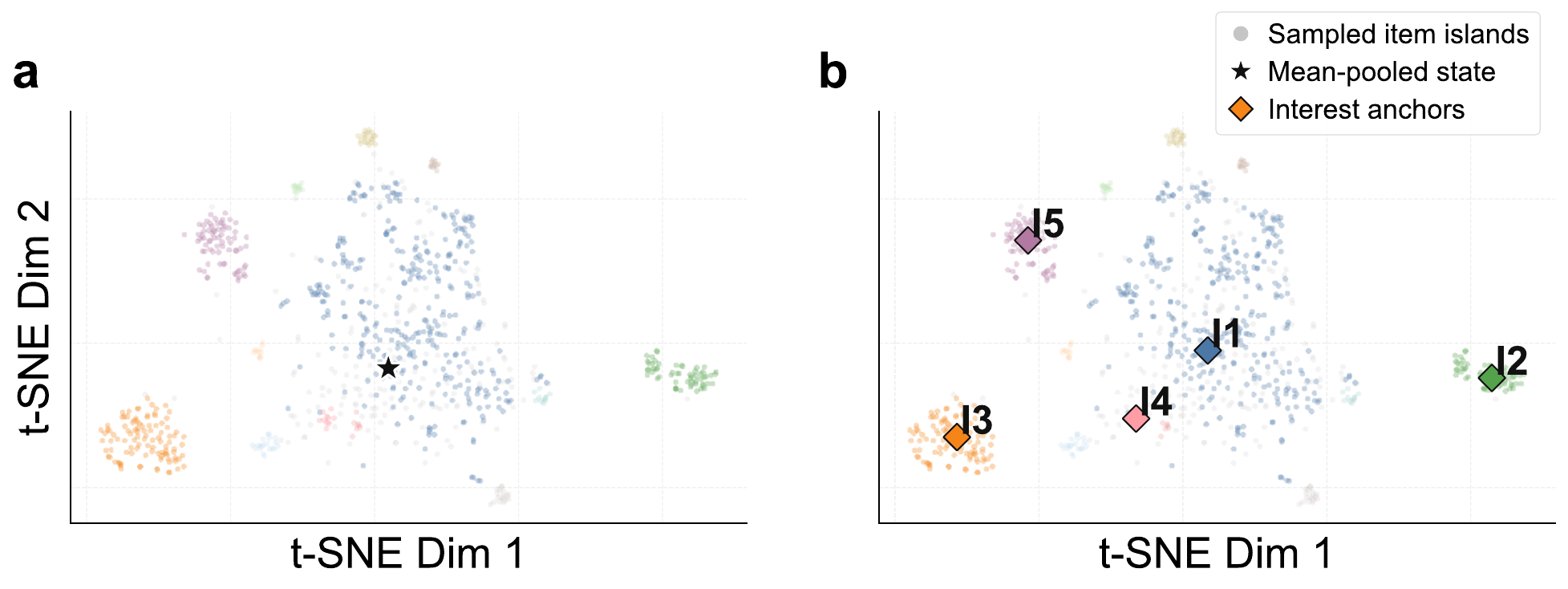}
    \caption{T-SNE visualization of user multi-interest behaviors on the Toys dataset.
      (a) A single averaged state suppresses minority interests;
      (b) multi-interest anchors preserve the semantic structure.}
    \label{fig:qualitative_tsne}
  \end{minipage}
  \hfill
  \begin{minipage}[t]{0.40\linewidth}
    \centering
    \includegraphics[width=\linewidth]{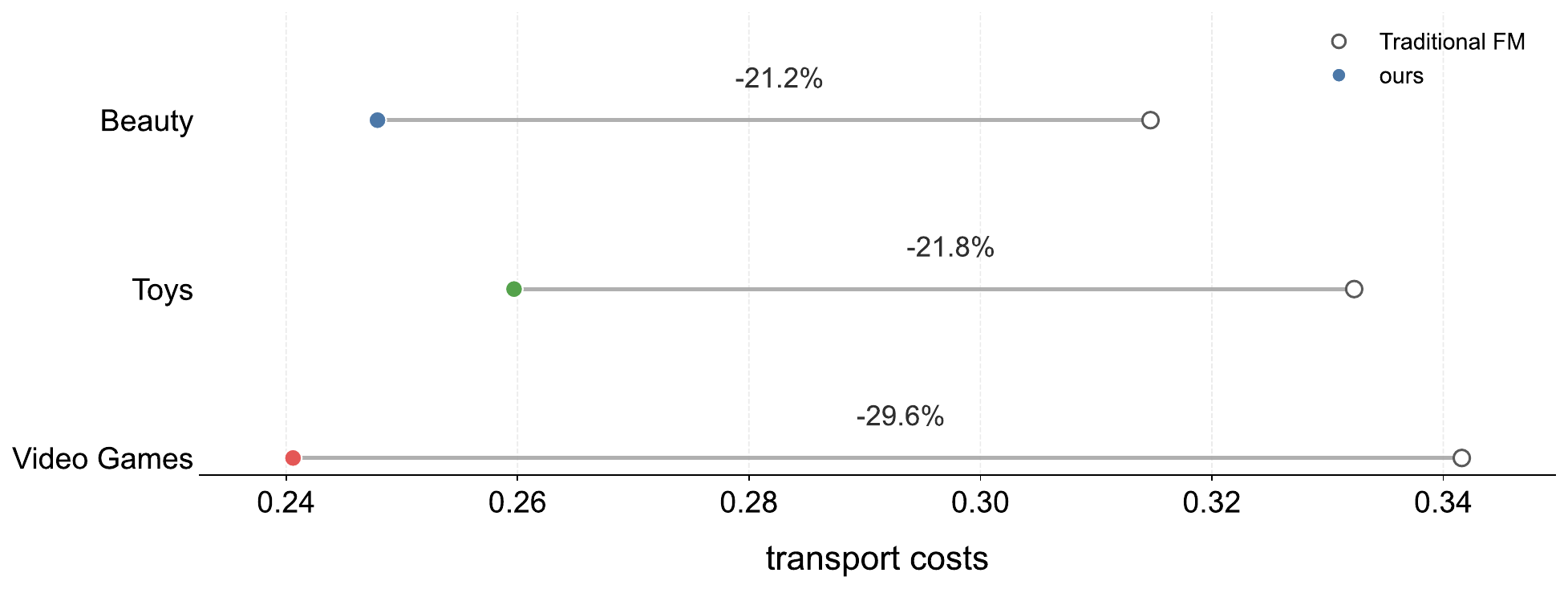}
    \caption{A comparison of transport costs between traditional flow matching and OT-FlowRec.}
    \label{fig:transport_costs}
  \end{minipage}

\end{figure}

Figure~\ref{fig:qualitative_tsne} shows the comparison between t-SNE projections \citep{ref64_vandermaaten2008_vdu} of historical item embeddings and the global mean-pooled state with the interest anchors learned by OT-FlowRec. A mean-pooled state represents these clusters with a single central point, whereas the learned anchors align with different clusters and retain the multi-peaked structure of concurrent user interests. This advantage arises from the MPSD model, which preserves interest-specific representations instead of averaging the entire interaction history into a single vector.

Beyond preserving the multi-peaked interest structure, Figure~\ref{fig:transport_costs} compares transport costs of OT-FlowRec with traditional flow matching based on random pairings. OT-FlowRec lowers the average transport cost by 24.2\% on three datasets, indicating better efficiency from source distribution to target preference prototypes. The improvement of OT-guided may be attributed to the coupling mechanism defined in Equation~\eqref{OT}, which jointly minimizes the trajectory costs of multiple bridges.

\section{Additional Implementation Details}
\label{app:additional_implementation_details}
We further tune OT-FlowRec on each dataset using validation Hit@10.
Unless otherwise specified, we fix the number of attention heads to 2, the feed-forward dimension to 512, the time-embedding dimension to 32.
The pre-extracted visual and textual features are kept frozen and are projected into the hidden space by two-layer MLPs.
The final item representation is obtained by a trainable softplus-gated fusion of ID, visual, and textual embeddings.
The user state is computed as the average of the last Transformer state and the mean-pooled valid sequence states.
The entropic OT coupling is solved by Sinkhorn iterations with regularization $\epsilon_{\mathrm{OT}}$, and the OT loss uses the expected squared transport cost under this coupling. For multi-prototype bridge inference, the top-$k$ prototypes are used as bridge conditions and their scores are aggregated by max pooling. We optimize softmax cross-entropy over all items.

The hyperparameter search space is summarized in Table~\ref{tab:search_space}.
Log-uniform parameters are sampled as $\exp(\mathcal{U}[a,b])$.
The dataset-specific best configurations selected on the validation set are reported in Table~\ref{tab:best_hyper}.
After hyperparameter selection, each final configuration is rerun with five random seeds $\{2022,2023,2024,2025,2026\}$, and we report the mean performance over these runs.

\begin{table}[H]
  \centering
  \caption{Hyperparameter search space of OT-FlowRec.}
  \label{tab:search_space}
  \small
  \begin{tabular}{ll}
    \toprule
    Hyperparameter                                      & Search space                   \\
    \midrule
    Learning rate                                       & $\exp(\mathcal{U}[-9.2,-6.0])$ \\
    Number of Transformer layers                        & $\{1,2,3\}$                    \\
    Hidden dimension $d$                                & $\{128,256\}$                  \\
    Hidden dropout / attention dropout                  & $\mathcal{U}[0.1,0.5]$         \\
    Number of prototypes $M$                            & $\{32,64,96,128\}$             \\
    ODE steps                                           & $\{2,4,6,8\}$                  \\
    Prototype temperature $\tau_{\mathrm{proto}}$       & $\exp(\mathcal{U}[-4.2,-1.9])$ \\
    Sinkhorn regularization $\epsilon_{\mathrm{OT}}$    & $\exp(\mathcal{U}[-4.0,-1.6])$ \\
    Sinkhorn iterations                                 & $\{5,10,20\}$                  \\
    Warm-up epochs                                      & validation-selected            \\
    $\lambda_{\mathrm{Proto}}$, $\lambda_{\mathrm{OT}}$ & $\exp(\mathcal{U}[-4.6,-0.7])$ \\
    \bottomrule
  \end{tabular}
\end{table}

\begin{table}[H]
  \centering
  \caption{Best hyperparameters of OT-FlowRec on each dataset.}
  \label{tab:best_hyper}
  \scriptsize
  \resizebox{\textwidth}{!}{
    \begin{tabular}{lccccccccccccc}
      \toprule
      Dataset      & LR                     & Layers & $d$ & Drop.   & Attn Drop. & $M$ & ODE & $\tau$   & $\epsilon_{\mathrm{OT}}$ & Iter. & Warm-up & $\lambda_{\mathrm{Proto}}$ & $\lambda_{\mathrm{OT}}$ \\
      \midrule
      Beauty       & $5.87{\times}10^{-4}$  & 1      & 128 & 0.39755 & 0.44607    & 96  & 6   & 0.03919  & 0.12053                  & 10    & 12      & 0.11967                    & 0.02177                 \\
      Toys         & $6.322{\times}10^{-4}$ & 1      & 256 & 0.45730 & 0.26735    & 96  & 6   & 0.02318  & 0.117279                 & 10    & 12      & 0.056853                   & 0.02742                 \\
      Video\_Games & $6.30{\times}10^{-4}$  & 2      & 128 & 0.49551 & 0.340265   & 96  & 6   & 0.084436 & 0.081655                 & 5     & 9       & 0.12009                    & 0.05440                 \\
      \bottomrule
    \end{tabular}}
\end{table}

\clearpage
\section{Additional Experimental Results}
\label{app:additional_experimental_results}
\begin{figure}[H]
  \centering
  \includegraphics[width=\columnwidth]{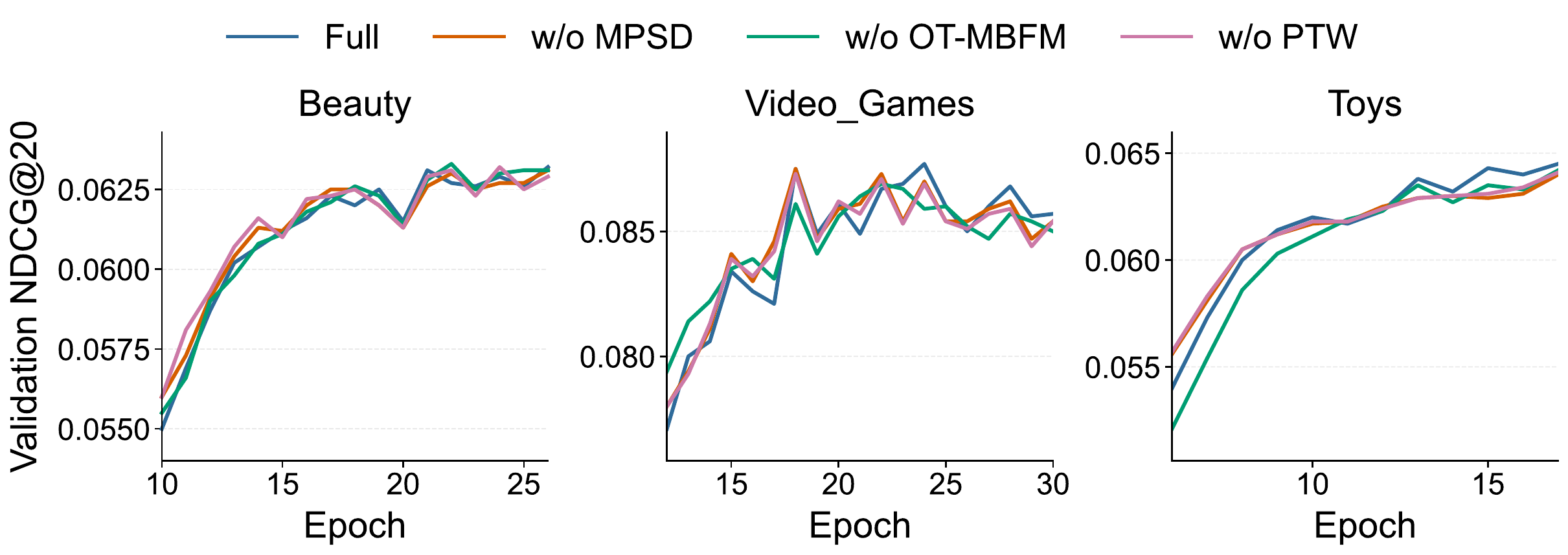}
  \caption{Validation N@20 curves of OT-FlowRec and the ablation variants}
  \label{fig:validation_curves}
\end{figure}

Validation curves corroborate these conclusions from the perspective of optimization dynamics. As Figure~\ref{fig:validation_curves} illustrates, the full model achieves more stable and higher validation N@20 scores across the Beauty, Video Games, and Toys datasets. On the Toys dataset, the full model converges faster during early training and maintains its lead in later stages, indicating that multi-interest modeling and optimal-transport bridge paths more effectively capture the evolution of user interests. Curves for the Video Games dataset exhibit more pronounced fluctuations, reflecting its more complex user interests and item semantics. While the full model maintains optimal performance in later stages, removing the OT-MBFM module causes frequent oscillation and insufficient convergence. This demonstrates that OT-MBFM not only improves final recommendation accuracy but also alleviates directional conflicts during training, yielding a more stable optimization trajectory. Although performance gaps among variants are smaller on the Beauty dataset, the full model still performs best. This suggests that the marginal benefits of individual modules diminish when user interest structures are smooth, whereas the advantages of the optimal-transport-guided multi-bridge flow mechanism become more prominent on datasets with diverse interests and complex semantics.

The training curve for the w/o PTW variant highlights the core contribution of PTW in aligning continuous distribution fitting with practical ranking metrics. By progressively adjusting the beta distribution bias for time-step sampling, PTW dynamically biases optimization weights toward the final states of the generation process, rather than relying on balanced full-path supervision. During early training stages, this variant still exhibits an upward trend comparable to the full model, suggesting that the basic optimal-transport bridge flow alone can learn effective directions for interest evolution. In later stages, however, it frequently plateaus or oscillates, failing to reliably achieve the optimal validation N@20 of the full model. This performance gap is especially evident on the Video Games and Toys datasets, where the full model maintains a distinct lead. These findings confirm that without dynamically biasing weights toward the final states of the evolution process, the model captures the overall transport path but struggles to refine the local discriminative capability required for practical top-k ranking.

\end{document}